\documentclass[aps,prl,twocolumn,floatfix,showpacs]{revtex4}

\usepackage{graphics}
\usepackage{graphicx}
\usepackage{amssymb}
\usepackage{amsfonts}
\usepackage{amsmath}
\usepackage{bm}

\DeclareMathOperator{\erf}{erf}

\begin{document}

\title{Exchange Coupling in a One-Dimensional Wigner Crystal}
\author{A. D. Klironomos}
\author{R. R. Ramazashvili}
\thanks{Present address: Max-Planck-Institut f\"{u}r Physik komplexer Systeme,
N\"{o}thnitzer Str. 38, 01187 Dresden, Germany}
\author{K. A. Matveev}
\thanks{On leave from Duke University, Durham, NC 27708-0305.}
\affiliation{Materials Science Division,
Argonne National Laboratory, Argonne, Illinois 60439, USA}
\date{\today}

\begin{abstract}

  We consider a long quantum wire at low electron densities.  In this
  strong interaction regime a Wigner crystal may form, in which electrons
  comprise an antiferromagnetic Heisenberg spin chain.  The coupling
  constant $J$ is exponentially small, as it originates from tunneling of
  two neighboring electrons through the segregating potential barrier.  We
  study this exponential dependence, properly accounting for the many-body
  effects and the finite width of the wire.

\end{abstract}

\pacs{73.21.Hb,75.10.Pq,75.30.Et,71.70.Gm}

\maketitle

Quantum wires exhibit a plethora of interesting phenomena.  In particular,
quantization of conductance, which is a fundamental manifestation of the
quantum nature of the electron, has been actively studied ever since its
first observation in quantum point contacts \cite{wees}.  The phenomenon
presents itself as very flat plateaus of linear conductance $G$ at integer
multiples of $G_0=2e^2/h$, as a function of gate voltage which tunes the
electron density in the wire.  Since the first observation of the
phenomenon, its various facets have been studied by measurements of
thermal transport \cite{molenkamp,wyss}, noise \cite{reznikov}, and
experiments on systems with superconducting elements \cite{beenakker}.

A new generation of experiments in quantum wires has revealed an
unexpected structure at low electron densities: a plateau at about
$0.7\,G_0$ for short
\cite{thomas,thomas1,cronenwett,kristensen1,kristensen} and at $0.5\,G_0$
for long quantum wires \cite{bkane,thomas2,reilly1,reilly2}.  These new
features have generated much interest as they are likely caused by
electron correlation effects.  The origin of the new plateau has not yet
been established; however, the experiments
\cite{thomas,thomas1,cronenwett} point to the important role played by the
electron spins.

In recent experiments of a different kind, involving tunneling between two
parallel ballistic quantum wires, mapping of the spectrum of spin and
charge excitations was achieved \cite{auslaender}.  In addition, unexpected
behavior indicating electron localization was observed at low densities.
Of particular interest is the concurrence of the localization with the
drops in the conductance steps, which indicates \cite{auslaender}
a possible connection with the 0.7 structure
\cite{thomas,thomas1,cronenwett,kristensen1,kristensen}.

Recent theoretical work suggests that qualitatively new transport
properties of quantum wires at low electron densities may be due to the
formation of a Wigner crystal state of electrons.  The latter is expected
to occur when the density is low enough for the potential energy of the
Coulomb repulsion to overwhelm the kinetic energy of the electrons in the
wire.  In the crystalline configuration, the electron spins form an
antiferromagnetic Heisenberg spin chain, with the exchange coupling
constant $J$ emerging as a new energy scale.  The Heisenberg exchange can
be viewed as arising from tunneling of two neighboring electrons through
the potential barrier created by their mutual repulsion, and by the
repulsion from all other electrons of the wire.  As a result, $J$ is
expected to be small compared to the Fermi energy $E_F$, and qualitatively
new transport properties of the quantum wires, such as the plateau at
$0.5\,G_0$, are expected \cite{matveev,matveev1} at temperatures in the
range $J\ll T\ll E_F$.  In addition to conductance measurements
\cite{bkane,thomas2,reilly1,reilly2}, the smallness of $J$ should have a
strong effect on the velocity of the spin excitations in the wire measured
in Ref.~\onlinecite{auslaender} and on transport properties of quantum
wires in the in-plane magnetic field.  Therefore it is important to have a
reliable estimate of $J$.

In this article we evaluate the exchange coupling constant $J$ in a
quantum wire at low electron density, when the electrons are in a Wigner
crystal state.  Similarly to the case of a two-dimensional crystal
\cite{roger}, the Coulomb interaction couples other electrons of the wire
to the tunneling of the exchanging pair, thus turning it into a many-body
process. We study this process, using a combination of analytic and
numerical tools in two distinct approximations.  First we consider a wire
of zero width, and subsequently account for the effect of a finite
width.

Disregarding the collective nature of the tunneling process, an estimate
of $J$ can be made \cite{hausler,matveev1} by considering a single pair
of exchanging electrons, with all the others held in fixed positions
separated by distance $b\equiv n^{-1}$.  The only dynamical variable in
this case is the relative coordinate $x$ of the two electrons.  Tunneling
through the potential barrier $U(x)$, created by their mutual repulsion
and by the repulsion from all the other electrons, lifts the ground state
degeneracy caused by the symmetry with respect to inversion $x\to -x$.
The exchange energy $J$ coincides with the exponentially small level
splitting.  In the semiclassical approximation it is given by
\begin{equation}
J=J^{*}\exp\left(-\frac{S_0}{\hbar}\right),
\quad S_0=\int_{-b}^b\!\!\mathrm{d}x\sqrt{mU(x)},
\label{WKB}
\end{equation}
where $m$ is the effective mass of the electrons.

Determining $J$ for a given density amounts to calculating the exponent
$S_0/\hbar$, with the prefactor $J^{*}$ being of secondary importance for
the problem at hand.  The action in Eq.~(\ref{WKB}) is easily brought to
the form $S_0=\hbar\eta_0/\sqrt{na_B}$, and the constant
$\eta_0\approx2.817$ is evaluated numerically \cite{matveev1}.  (Here
$\epsilon$ is the dielectric constant, and $a_B=\epsilon\hbar^2/me^2$ is
the effective Bohr radius of the semiconductor host).  The precision of
this estimate is unclear, as only the two exchanging electrons were
allowed to move.

To account for the motion of all the electrons, we treat the tunneling
using the instanton method, in which the action is represented as an
imaginary-time integral.  By measuring distance and time in units of
$b$ and $(\epsilon m b^3/e^2)^{1/2}$, respectively, we find the
dimensionless action
\begin{equation}
\eta[\{X_j(\tau)\}]=\int_{-\infty}^\infty\!\!\mathrm{d}\tau
     \!\left[\sum_{j}
     \frac{\dot{X}_j^2}{2}
    +\sum_{j<i}\frac{1}{|X_j(\tau)-X_i(\tau)|}
  \right]\!\!.
 \label{eq:action}
\end{equation}
Here $X_j$ is the dimensionless coordinate of the $j$-th electron.  In
this approach the exchange constant is given by
\begin{equation}
  \label{eq:J}
  J=J^*\exp\left(-\frac{\eta}{\sqrt{na_B}}\right),
\end{equation}
where $\eta$ is the action (\ref{eq:action}) minimized over the trajectories
$X_j(\tau)$ of all the particles.

The approximation (\ref{WKB}) is equivalent to imposing a constraint
$X_j(\tau)\equiv j$ for all $j\neq 0,$ 1.  By releasing this constraint we
minimize the action over more variables.  Thus allowing for the
participation of all the electrons in the exchange process will result in
$\eta<\eta_0$.

In the course of the tunneling, near neighbors of the exchanging pair
undergo displacements comparable to the inter-electron distance.  The
contribution of these electrons to the tunneling action $S_0$ can only
be studied by numerical means.  By contrast, the displacements of remote
electrons are small and vary smoothly, allowing for a continuous description
and analytic treatment of their contribution to $\eta$.  Combining the
two contributions permits a reliable evaluation of $\eta$.

We study the long distance contribution first. To this end, we modify the
problem as follows.  Starting with the nearest neighbors of the exchanging
pair, a large number $l\gg1$ of electrons are assumed to be constrained to
their equilibrium positions.  As a result, all the moving neighbors are
far from the exchanging pair, which enables analytic treatment of the
problem.  Furthermore, it will be convenient to assume that only $N\gg l$
of the subsequent electrons can move.  Apart from taking the limit $N\to
\infty$, this assumption will enable us to study how this limit is
approached at large $N$.

We treat the $N$ mobile electrons as a continuous string described by its
displacement $u(X,\tau)$ from equilibrium, $l<X<l+N$.  Since $l\gg1$, the
string is far from the exchanging pair, and the resulting displacement is
small, $u\ll1$.  This enables us to account for the coupling of the string
to the exchanging pair in linear order in $u$ and present the string
contribution $\eta_s$ to the action (\ref{eq:action}) as
\begin{equation}
\label{Ss}
\eta_s=\int\!\frac{\mathrm{d}\omega}{2\pi} \sum_{k}
       \left\{(\omega^2+\omega_k^2)|u_{k\omega}|^2
          + 12 f_{\omega}u_{k\omega}^{*}\left[\frac{1}{X^4}\right]_{k}
       \right\},
\end{equation}
Here the parameter $k$ labels the (normalized) eigenmodes $\Psi_k(X)$ of the
string, $\omega_k$ is the respective plasmon frequency, and $[X^{-4}]_k$
denotes the expansion coefficient of the function $X^{-4}$ in the eigenmode
basis.  The coupling of the string to the exchanging pair is accounted for
by the function $f_\omega$ defined as the Fourier transform of
$f(\tau)=X_0(\tau)X_1(\tau)$.

According to Eq.~(\ref{Ss}) the coupling of a remote mobile electron to
the exchanging pair decays as $1/X^4$.  This can be understood by noticing
that symmetric displacement of the exchanging electrons, $X_1=1-X_0$,
creates a quadrupole potential $\propto 1/X^3$ at large distances.  To
linear order in $u(X)$, the displacement of a mobile electron is
equivalent to creation of a dipole at point $X$, whose interaction with
the quadrupole field decays as $1/X^4$.  Since $X>l$, the coupling of the
string to the exchanging pair is weak at $l\gg1$.  The feedback
effect of the string upon the exchanging pair is further reduced by the
same parameter.  Thus we neglect the latter effect and substitute into
Eq.~(\ref{eq:action}) the function $f_\omega$ computed with unperturbed
functions $X_0(\tau)$ and $X_1(\tau)$.

We then minimize the action~(\ref{Ss}), and find the string contribution
to the total action
\begin{equation}
\label{action}
  \eta_s = -\frac{9|f_{0}|^{2}}{2}
            \sum_{k} \frac{1}{\omega_k}
            \left[ \int_{l}^{l+N}\!\!\mathrm{d}X
                  \frac{\Psi_{k}(X)}{X^4} \right]^2.
\end{equation}
Here we used the fact that at frequencies well below the Debye frequency
of plasmons the Fourier transform $f_\omega$ of $f(\tau)$ smoothly
approaches its zero-frequency limit $f_0$.

The eigenmodes $\Psi_k(X)$ are essentially standing plasmon waves in the
string with wave number $k$; only near the edges does their shape slightly
deviate from the sinusoidal one.  Using the proper plasmon dispersion
\begin{equation}
\omega_k=k \sqrt{2\ln k^{-1}},
\end{equation}
we find the following asymptotic behavior in the limit of large $N$:
\begin{equation}
\label{asymp}
\eta_N=\eta + \frac{\alpha}{N^2\sqrt{\ln N}}.
\end{equation}
The expressions for the two constants $\eta$ and $\alpha$ involve
parameters $f_0$ and $\Psi'_k(l)$ which are easily found numerically.  As
a result we obtain
\begin{equation}
  \label{eq:parameters}
  \eta = \eta_0 - \frac{0.020}{l^6\sqrt{\ln l}},
  \quad
  \alpha = \frac{0.011}{l^4}.
\end{equation}

In the limit of large number of mobile particles $N\to\infty$, the
numerator of the exponent in the expression (\ref{eq:J}) for the exchange
constant takes the limiting value $\eta$; as expected, $\eta<\eta_0$.
Somewhat surprisingly, however, the total effect is numerically small:
even at $l\sim 1$ we find $\eta_0-\eta\sim 0.02$, i.e., the total
correction is of the order of one percent of the unperturbed value
$\eta_0\approx 2.817$.

\begin{figure}[!t]
\includegraphics[height=6.0cm,clip]{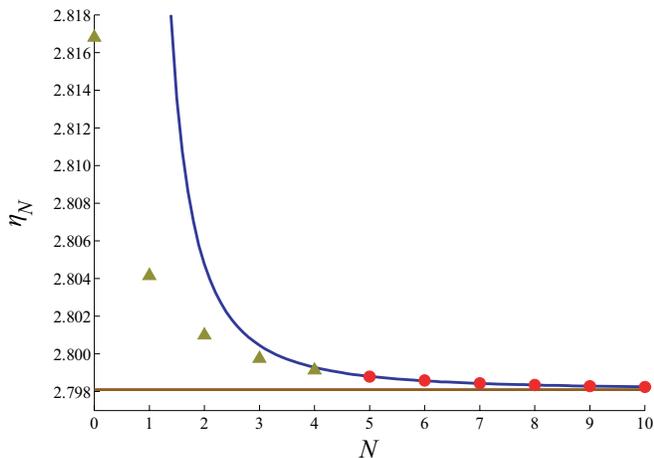}
\caption{Triangles and circles represent the calculated exponents.  Only
  the values represented by the circles ($N=6\rightarrow 11)$ were used to
  fit to the analytically predicted asymptotic behavior (\ref{asymp}).
  The horizontal line indicates the value of the exponent in the limit
  when all electrons participate in the tunneling.  For this particular
  fit we obtained $\eta=2.7981$.}
\label{fig_1}
\end{figure}

It is important to note that the above approach relies on the large
parameter $l\gg1$, which is the number of immobile electrons between the
mobile ones and the exchanging pair.  In the actual exchange process, all
electrons move and contribute to the action.  Fortunately, the
contribution to the tunneling action from a few near neighbors of the
exchanging pair can be found numerically.  To this end, we minimize the
action (\ref{eq:action}) with a finite number $N$ of mobile electrons on
each side of the exchanging pair, while keeping the remaining electrons at
their equilibrium positions.  Mathematically, this procedure is equivalent
to solving a system of $N$ second order differential equations of motion
with the boundary conditions $X_1(-\infty)=1$, $X_1(+\infty)=0$, and
$X_j(-\infty)=X_j(+\infty) = j$ for $j=2$, 3, \ldots, $N+1$.  In practice,
we were able to perform this calculation for up to $N=10$; the results for
the respective action $\eta_N$ are shown in Fig.~\ref{fig_1}.  The
calculated asymptotic behavior (\ref{asymp}) can be used to fit these
results.  For instance, using the last six calculated values
$\eta_{5},\dots,\eta_{10}$, we find an excellent fit with fitting
parameters $\eta=2.7981$, and $\alpha=0.0221$.  In order to obtain a more
accurate result and estimate uncertainty bounds, we considered a series of
two different kinds of fitting schemes. One consisted of fits using
various numbers of $\eta_N$ points; for the other, we introduced a third
fitting parameter, namely an offset to $N$.  Note that the latter does not
contradict the predicted asymptotic behavior (\ref{asymp}).
  By combining the analytical treatment with numerical calculation in this
fashion we have obtained the exponent $\eta=2.79805\pm0.00005$ in
Eq.~(\ref{eq:J}) in the limit when all the electrons participate in the
exchange process.

Since the effect of neighboring electrons upon $\eta$ in Eq.~(\ref{eq:J})
is fairly weak, it is important to consider other corrections which may
potentially affect the exchange coupling in quantum wires.  The most
important correction is due to the fact that quantum wires have finite
width $w$.  As a result, the exchanging electrons can utilize the
dimensions transverse to the wire axis to minimize the Coulomb repulsion
at short separations.  Therefore, the short distance behavior of the
interaction is strongly affected by the regularization of the singularity
inherent in the previous one-dimensional idealization.  The implication is
that the action should be further reduced.

There are two limits for which we can calculate the finite-width
corrections to the action. The small width limit, $w\ll a_B$, implies that
the typical electron interaction energy $e^2/\epsilon w$ is much smaller than
the subband spacing in the wire $\hbar^2/mw^2$.  Then the processes of
electron scattering to higher subbands are negligible, and the effect of
the finite width of the wire amounts to small smearing of electron density
in transverse direction.  Thus we calculate an effective one-dimensional
interaction $U_w(x)$ as an average of the three-dimensional Coulomb
interaction over the transverse coordinates, using the ground state wave
function in the confining potential.  This procedure changes the
interaction potential in Eq.~(\ref{WKB}) only at distances $|x|\lesssim
w$, which enables us to estimate the correction to the action $S_0$ as
\begin{equation}
\label{corr_1}
\Delta
S=\int_{-b}^b\!\!\!\mathrm{d}x\left(\sqrt{mU_w(x)}-\sqrt{mU(x)}\right)
 =-\hbar\kappa\sqrt\frac{w}{a_B},
\end{equation}
where only the numerical prefactor $\kappa$ is sensitive to the details of
the confining geometry.  As an illustration, we consider an axially
symmetric quadratic confining potential and define $w$ as root-mean-square
deviation of the electron from the axis in the ground state.  At short
distances $|x|\ll b$, we find an effective interaction
\begin{equation}
U_w(x)=\frac{e^2\sqrt{\pi}}{2w\epsilon}e^{(x/2w)^2}
       \left[1-\erf\left(\frac{|x|}{2w}\right)\right],
\end{equation}
which, of course, coincides with $e^2/\epsilon|x|$ at $|x|\gg w$.
Substituting this expression into Eq.~(\ref{corr_1}), we find
$\kappa\approx2.99$.  We estimated the corrections to $\Delta S$ from
small admixture of higher subbands to be of order $(w/a_B)^{5/2}$.
\begin{figure}[!t]
\includegraphics[height=3.3cm,clip]{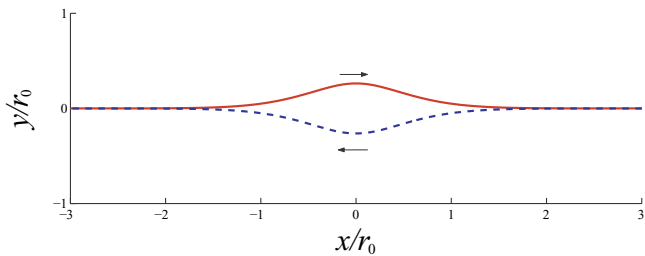}
\caption{The classical exchange trajectories obtained by solving
  numerically the equations of motion for two electrons (zero density
  limit) in a wire of width $w\gg a_B$ with a quadratic confining
  potential.  Arrows indicate the direction of motion.  The unit of length
  is $r_0=2a_B(w/a_B)^{4/3}$.}\
\label{instanton}
\end{figure}

In the large width limit, $w\gg a_B$, we can once again consider a
semiclassical approximation analogous to the one-dimensional one we
employed previously.  Indeed, in the instanton approach the trajectories
of two exchanging electrons behave as shown in Fig.~\ref{instanton}.  The
electrons pass each other at a distance $r_0$, at which the Coulomb
repulsion is of the order of the confining potential.  For example, in the
case of quadratic potential we estimate $e^2/\epsilon r_0\sim
(\hbar^2/mw^2) (r_0/w)^2$ and find $r_0\sim (w^4/a_B)^{1/3}$.  Since $r_0\gg
w$, the quantum particles always stay close to the instanton trajectory,
and the semiclassical approach is applicable.  Furthermore, since the two
electrons pass each other at distance of order $r_0$, their interaction
potential is regularized at that length scale.  This reduces the tunneling
action in a way similar to the small width case (\ref{corr_1}), except
that instead of $w$ one should substitute $r_0$.  Thus the
correction to the exchange action (\ref{WKB}) is
\begin{equation}
\label{corr_2}
\Delta S=-\hbar\tilde{\kappa}\left(\frac{w}{a_B}\right)^{2/3}
\end{equation}
at $w\gg a_B$.  The correction grows faster as a function of $w$ than in
the regime $w\ll a_B$.  This generic feature of the problem occurs for any
shape of the confining potential (other than the hard-wall one).  However,
the 2/3 power-law dependence is valid only for quadratic potential.
\begin{figure}[!t]
\includegraphics[height=6.25cm,clip]{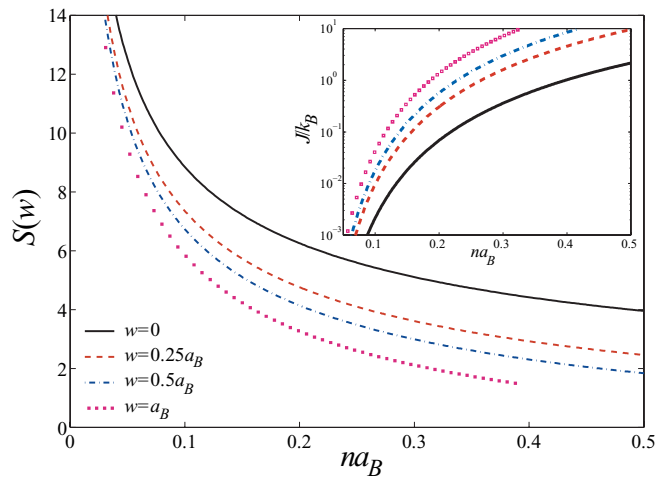}
\caption{The calculated action as a function of density for various wire
  widths $w$.  Inset: the exchange energy $J$ in Kelvins as a function
  of density for the same set of wire widths as in the main figure, using
  the estimate of the prefactor obtained in Ref.~\onlinecite{matveev1} and
  parameters pertaining to GaAs.}
\label{plot_action}
\end{figure}

To find the coefficient $\tilde\kappa$, we compute the instanton for the
exchange of two particles numerically by solving the classical equations
of motion in the inverted potential.  In the low-density regime, $n\ll
r_0^{-1}$, the trajectories deviate from straight lines only at short
distances $\sim r_0$, Fig.~\ref{instanton}.  We thus evaluate the
correction to $S_0$, by finding the difference of the two corresponding
actions calculated up to a large distance cutoff.  This calculation gives
$\tilde{\kappa}\approx 2.03$.

We are now in a position to calculate the exchange constant as a function
of density for both the one-dimensional wire, and wires of various
representative widths.  The two corrections to the action,
Eqs.~(\ref{corr_1}) and (\ref{corr_2}), have comparable magnitudes at
$w\simeq11a_B$.  Considering that for typical confinement geometries in
realistic devices $w$ does not exceed $a_B$, we use the finite-width
correction in the form (\ref{corr_1}).  In Fig.~\ref{plot_action} we show
the corresponding curves as a function of density.

Summarizing, we have found the exchange energy in a Wigner crystal of
electrons in a quantum wire in the exponential approximation.  The
tunneling process involves many electrons neighboring the exchanging
pair.  Their effect was accounted for by combining a numerical
calculation for the near neighbors of the exchanging pair with an
analytic calculation of the asymptotic behavior for the distant
ones.  Finally, we found the substantial corrections due to a finite
width of the wire in both small- and large-width limits.

Work supported by the U. S. Department of Energy, Office of
Science, under Contract No. W-31-109-ENG-38.


\begin{thebibliography}{99}
\bibitem{wees} B.~J. van Wees \textit{et al.}, Phys.\ Rev.\ Lett.
  \textbf{60}, 848 (1988); D.~A. Wharam \textit{et al.}, J. Phys. C
  \textbf{21}, L209 (1988).
\bibitem{molenkamp} L.~W. Molenkamp \textit{et al.}, Phys.\ Rev.\ Lett.
  \textbf{65}, 1052 (1990); P. St\v{r}eda, J.\ Phys.\ Cond.\ Matter
  \textbf{1}, 1025 (1989).
\bibitem{wyss} R.~A. Wyss \textit{et al.}, Appl.\ Phys.\ Lett.
  \textbf{66}, 1144 (1995).
\bibitem{reznikov} M. Reznikov \textit{et al.}, Phys.\ Rev.\ Lett.
  \textbf{75}, 3340 (1995).
\bibitem{beenakker} C.~W.~J.~Beenakker, H. van Houten, Phys.\ Rev.\ Lett.
  \textbf{66}, 3056 (1991); A. Furusaki, H. Takayanagi, M. Tsukada, Phys.\
  Rev.\ B \textbf{45}, 10563 (1992); H. Takayanagi, T. Akazaki, J. Nitta,
  Phys.\ Rev.\ Lett. \textbf{75}, 3533 (1995).
\bibitem{thomas} K.~J. Thomas \textit{et al.}, Phys.\ Rev.\ Lett.
  \textbf{77}, 135 (1996).
\bibitem{thomas1} K.~J. Thomas \textit{et al.}, Phys.\ Rev.\ B
  \textbf{58}, 4846 (1998).
\bibitem{cronenwett} S. Cronenwett \textit{et al.}, Phys.\ Rev.\ Lett.
  \textbf{88}, 226805 (2002).
\bibitem{kristensen1} A.~Kristensen \textit{et al.}, J.\ Appl.\ Phys.
  \textbf{83}, 607 (1998).
\bibitem{kristensen} A.~Kristensen \textit{et al.}, Phys.\ Rev.\ B
  \textbf{62}, 10950 (2000).
\bibitem{bkane} B.~E. Kane \textit{et al.}, Appl.\ Phys.\ Lett.
  \textbf{72}, 3506 (1998).
\bibitem{thomas2} K.~J. Thomas \textit{et al.}, Phys.\ Rev.\ B
  \textbf{61}, R13365 (2000).
\bibitem{reilly1} D.~J. Reilly \textit{et al.}, Phys.\ Rev.\ B
  \textbf{63}, 121311(R) (2001).
\bibitem{reilly2} D.~J. Reilly \textit{et al.}, Phys.\ Rev.\ Lett.
  \textbf{89}, 246801 (2002).
\bibitem{auslaender} O.~M. Auslaender \textit{et al.}, Science
  \textbf{295}, 825 (2002); O.~M. Auslaender \textit{et al.}, unpublished
  (2005).
\bibitem{matveev} K.~A. Matveev, Phys.\ Rev.\ Lett. \textbf{92}, 106801
  (2004).
\bibitem{matveev1} K.~A. Matveev, Phys.\ Rev.\ B \textbf{70}, 245319
  (2004).
\bibitem{roger} See, e.g., M. Roger, Phys.\ Rev.\ B \textbf{30}, 6432
  (1984).
\bibitem{hausler} W.~H\"{a}usler, Z. Phys. B \textbf{99}, 551 (1996).
\end{thebibliography}
\end{document}